\documentclass[prb,twocolumn,showpacs,noshowkeys,
preprintnumbers,amsmath,amssymb]{revtex4}

\usepackage{graphicx}
\usepackage{dcolumn}
\usepackage{bm}
\usepackage{hyperref}%

\begin{document}

\preprint{}

\title{$f(\alpha)$ Multifractal spectrum at strong and weak disorder}
\author{E. Cuevas}
\email{ecr@um.es}
\homepage{http://bohr.fcu.um.es/miembros/ecr/}
\affiliation{Departamento de F{\'\i}sica, Universidad de Murcia,
E-30071 Murcia, Spain.}

\date{\today}

\begin{abstract}
The system size dependence of the multifractal spectrum $f(\alpha)$
and its singularity strength $\alpha$ is investigated numerically.
We focus on one-dimensional (1D) and 2D disordered systems with
long-range random hopping amplitudes in both the strong and the weak
disorder regime. At the macroscopic limit, it is shown that $f(\alpha)$
is parabolic in the weak disorder regime. In the case of strong
disorder, on the other hand, $f(\alpha)$ strongly deviates from
parabolicity. Within our numerical uncertainties it has been found
that all corrections to the parabolic form vanish at some finite value
of the coupling strength.
\end{abstract}



\maketitle

\section{Introduction}

In recent years, it has become clear that the eigenfunctions
of noninteracting disordered systems at the critical point of
a metal-insulator transition (MIT) have multifractal scaling
properties. Hence, multifractal analysis has become a standard
tool for treating the strong wave-function fluctuations which
arise near criticality. \cite{FE95} One way to characterize
wave-function statistics is through the corresponding multifractal
$f(\alpha)$ spectrum. This spectrum of the local density is
universal at the critical point, and the exponents such as
$\alpha_0$, which describes the scaling behavior of the typical
local electron density, are suitable for describing the phase
transition. \cite{Ja94} Finding the true $f(\alpha)$ curve is
essential in order to fully understand MITs.

MITs depend on the dimensionality and symmetries
of the system and can occur in both the strong disorder and the
weak disorder regime (strong-coupling or weak-coupling regime,
respectively, in the language of field theory). Each regime is
characterized by the corresponding coupling strength that
depends on the ratio between diagonal disorder and the off-diagonal
transition matrix elements of the Hamiltonian. \cite{Ef83} We wish
to point out that most of the already published analytical results
concerning the $f(\alpha)$ spectrum, which we summarize in the next
paragraphs, have been obtained in the weak disorder limit
(weak-coupling approach).

Using the renormalization group in $d=2+\epsilon$ dimensions 
($\epsilon=\alpha_0-d \ll 1$), it was found \cite{CP86} that
the singularity spectrum is parabolic,
\begin{equation}
f(\alpha)=d-\frac {(\alpha-\alpha_0)^2}{4(\alpha_0-d)}\;,
\label{parap}
\end{equation}
up to $O(\epsilon^2)$. This equation was justified for the case
of unbroken time-reversal symmetry (orthogonal universality class)
in the weak-coupling limit.

Recently, in Ref. 5 it was found that a parabolic spectrum
(PS) is fulfilled for the critical wave function of a
two-dimensional (2D) Dirac fermion in a random vector potential
(chiral universality class). These results were obtained by
mapping the problem to a Gaussian field theory in an ultrametric
space (a Cayley tree). For a deep understanding of the origin of
the multifractal behavior in this model, see also Ref. 6.

The field-theoretical approaches in the context of the integer quantum
Hall transition (unitary universality class) has been addressed more
recently. Based on the application of conformal field theory to the
critical point of a 2D Euclidean field theory,
Bhaseen \textit{et al.} \cite{BK00} conjectured that the singularity
spectrum is exactly parabolic at the plateau transitions. Using the
same general principles of conformal field theory, \cite{Zi99,JM99}
a PS was conjectured for a closely related quantity, the two-point
conductance.

Given the present status of this problem, it would be highly desirable
to check (i) whether the PS is fulfilled at weak disorder strength and
(ii) whether or not the PS holds in the strong disorder regime, where it
is known that all realistic MITs, such as the conventional Anderson
transition in 3D, take place. Unfortunately, because of the absence of
a small parameter, no quantitative predictions for the critical properties
and, in particular, for the $f(\alpha)$ spectrum have been made in the
strong-coupling regime. In the absence of such predictions, the problem
must be addressed using numerical calculations. That is why we are
interested in exploring the role of the disorder strength on the $f(\alpha)$
spectrum. Specifically, we want to analyze how the singularity spectrum
evolves in the face of increasing coupling strength, i.e., when
the transition shifts from the weak disorder regime to the strong disorder
regime. 

This paper is organized as follows. In Sec. I, we introduce
the model and the methods used for the calculations. The results 
for the multifractal spectrum in 1D and 2D are shown in Secs.
II A and II B, respectively. Finally, the conclusions are presented
in Sec. III.

\section{Model and methods}

For computation of the multifractal spectrum, we use the
standard box-counting procedure, \cite{Ja94} first dividing
the system of $L^d$ sites into $N_l=(L/l)^d$ boxes of linear
size $l$ and determining the box probability of the wave function
in the $i$ box by $\mu_i(l)=\sum_n |\psi_{kn}|^2$, where the 
summation is restricted to sites within that box and $\psi_{kn}$
denotes the amplitude of an eigenstate with energy $E_k$ at site
$n$. The normalized $q$th moments of this probability
$\mu_i(q,l)=\mu_i^q(l)/\sum_{j=1}^{N_l} \mu_j^q(l)$ constitute
a measure. From this, the Lipschitz-H\"{o}lder exponent
or singularity strength can be obtained, \cite{CJ89}
\begin{equation}
\alpha_q(L)=\lim_{\delta \to 0} \frac {\sum_{i=1}^{N_l}
\mu_i(q,l)\ln \mu_i(1,l)}{\ln \delta}\;,\label{alfadeq}
\end{equation}
as well as the corresponding fractal dimension
\begin{equation}
f(\alpha_q(L))=\lim_{\delta \to 0} \frac {\sum_{i=1}^{N_l}
\mu_i(q,l)\ln \mu_i(q,l)}{\ln \delta}\;,\label{fdeq}
\end{equation}
which yields the characteristic singularity spectrum $f(\alpha)$
in a parametric representation. In Eqs. (\ref{alfadeq}) and
(\ref{fdeq}), $\delta=l/L$ denotes the ratio of the box sizes and
the system size. If the $q$th moments of the measure counted in
all boxes are proportional to a power $\tau_q$ of the box size,
$\langle \sum_{i=1}^{N_l} \mu_i^q(l) \rangle
\propto l^{-\tau_q}$, multifractal behavior may be derived.
The $f(\alpha)$ spectrum and the mass exponent $\tau_q$ are related
by a symmetric Legendre transformation, $f(\alpha)=\alpha q-\tau_q$,
with $\alpha=d\tau_q/dq$ and $q=df(\alpha)/d\alpha$. From Eq.
(\ref{parap}), one can easily obtain
\begin{equation}
\alpha_0-d=d-\alpha_1\;. \label{a0a1}
\end{equation}
In what follows, we will first use this relation to check whether
or not the PS is valid in each regime studied. First, $\alpha_q(L)$
and $f(\alpha_q(L))$ were calculated for different system sizes
and then extrapolated to the macroscopic limit
$\alpha_q=\lim_{L \to \infty} \alpha_q(L)$ and
$f(\alpha_q)=\lim_{L \to \infty} f(\alpha_q(L))$. We wish to
clarify that the calculation of $\alpha_q(L)$ and $f(\alpha_q(L))$
is suitable only if the conditions \cite{Ja94}
\begin{equation}
a\ll l < L \ll \xi \label{ineq}
\end{equation}
are satisfied, where $\xi$ is the localization or correlation length
and $a$ is the lattice spacing (or any microscopic length scale of
the system, such as the cyclotron radius).

Usually, the disorder-induced MIT is investigated for Hamiltonians
with short-range, off-diagonal matrix elements (e.g., the
canonical Anderson model). Another class of Hamiltonian exhibiting
an MIT in arbitrary dimension $d$ is formed by those that include
long-range hopping terms. The effect of long-range hopping on
localization was originally considered by Anderson \cite{An58} for
randomly distributed impurities in $d$ dimensions with the
$V(\bm{r}-\bm{r'}) \sim |\bm{r}-\bm{r'}|^{-\beta}$
hopping interaction. It is known \cite{An58,Le89} that all states are
extended for $\beta \le d$, whereas for $\beta > d$, the states are
localized. Thus, one can study the MIT by varying the exponent $\beta$
at fixed disorder strength.
At the transition line $\beta=d$, a real-space renormalization group
can be constructed for the distribution of couplings. \cite{Le89,Le99}
These models are most convenient for studying critical properties
because the exact critical point is known ($\beta=d$) and they allows us to
treat the 1D and 2D cases, thus reaching larger system sizes and reducing
the numerical effort. Here, we will concentrate on the 1D and 2D versions
of these models with orthogonal symmetry.
In our calculations, we consider a small energy window, containing about
8\% of the states around the center of the spectral band. The number
of random realizations is such that the number of critical states included
for each $L$ is roughly $4\times 10^5$, while in order to reduce the edge
effects, periodic boundary conditions in all directions are included.
Using methods based on level statistics, we checked that the normalized
nearest level variances \cite{Cu99} are indeed scale invariant at each
critical point studied.

\subsection{1D system}

First, we concentrate on the intensively studied power-law random
banded matrix model (PRBM). \cite{Mi00,ME00,CG01,CO02,Va02} The
corresponding Hamiltonian that describes
a disordered 1D sample with random long-range hopping is represented
by real symmetric matrices, whose entries are randomly drawn from a
normal distribution with zero mean,
$\left\langle {\cal H}_{ij} \right\rangle =0$, and a variance which
depends on the distance between lattice sites,
\begin{equation}
\left\langle |{\cal H}_{ij}|^2\right\rangle =\frac{1}{1+(|i-j|/b)^2}
\times\left\{\begin{array}{ll}
                    \tfrac{1}{2} \ ,\quad & i\neq j\\
                    1 \;\;\;\,\ ,\quad & i=j\;.
       \end{array}\right.
\label{h1dor}
\end{equation}
The model describes a whole family of critical theories parametrized
by $0<b<\infty$, which determines the critical dimensionless conductance
in the same way as the dimensionality labels the different Anderson
transitions. In the two limiting cases $b \gg 1$  and $b \ll 1$,
which correspond to the weak and the strong disorder limits,
respectively, some critical properties have been derived analytically.
\cite{Mi00,ME00,KM97,MF96,KT00} The system size ranges between $L=40$
and $L=1800$, and $0.03 \le b \le 30$. We restrict ourselves to values
of $q \ge 0$, which correspond to $\alpha \le \alpha_0$.

Using the exact eigenstates of Eq. (\ref{h1dor}) from numerical
diagonalizations, we obtained the size dependence of the scaling
exponents $\alpha_q(L)$ and
$f(\alpha_q(L))$ at the critical point of the finite system. In order to
extrapolate to macroscopic systems, we propose a finite-size correction
to $\alpha_q(L)=\alpha_q+a_qL^{-y_q}$, where the irrelevant exponent
$y_q > 0$. For all values of $q$, we found that the exponent $y_q$ hardly
differs from unity. The same behavior is found for $f(\alpha_q(L))$. Thus,
we can write
\begin{equation}
\alpha_q(L)=\alpha_q+a_q/L \;, \quad
f(\alpha_q(L))=f(\alpha_q)+b_q/L\;,\label{s2l}
\end{equation}
with $\alpha_q$, $a_q$, $f(\alpha_q)$, and $b_q$ being adjustable
parameters. As we will see below, our numerical data strongly support
this behavior.

As a first step, for each value of $q$, we evaluate the numerators on the
right-hand sides of Eqs. (\ref{alfadeq}) and (\ref{fdeq}), respectively,
for decreasing box sizes, and we calculate $\alpha_q(L)$ and $f(\alpha_q(L))$
from the slopes of the graphs of the numerators versus $\ln \delta$. In order
to satisfy the inequalities of Eq. (\ref{ineq}), we take $\delta$ to be in the
interval $(0.1,0.5)$. Figure 1 provides an example of the linear fit to
$\sum_i \mu_i(q,l)\ln \mu_i(1,l)$ versus $\ln \delta$ for two values
of $q$, $q=0$ (open symbols, right axis) and $q=1$ (solid symbols, left
axis), and different system sizes: $L=40$ (circles), 60 (triangles), 120
(diamonds), and 1800 (squares). Clearly, there is no ambiguity in the
determination of the slopes that correspond to the values of $\alpha_0(L)$
and $\alpha_1(L)$. These slopes, which are summarized in Fig. 2, have been
obtained for $b=0.3$.
\begin{figure}
\begin{center}
\includegraphics[width=8.0cm]{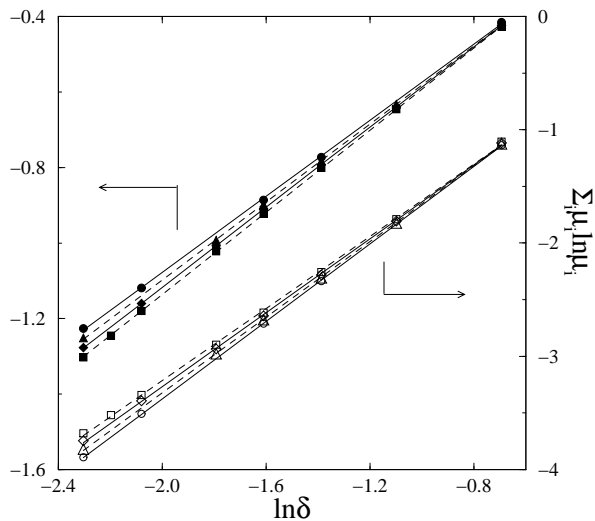}
\caption{\label{fig1} 
$\sum_i \mu_i(q,l)\ln \mu_i(1,l)$ as a function of $\ln \delta$ for
$q=0$ (open symbols, right axis) and $q=1$ (solid symbols, left axis)
and different system sizes: $L=40$ (circles), 60 (triangles), 120
(diamonds), and 1800 (squares). The straight lines whose slopes
correspond to the values of $\alpha_0(L)$ and $\alpha_1(L)$ are linear
fits to Eq. (\ref{alfadeq}).}
\end{center}
\end{figure}
\begin{figure}
\begin{center}
\includegraphics[width=8.0cm]{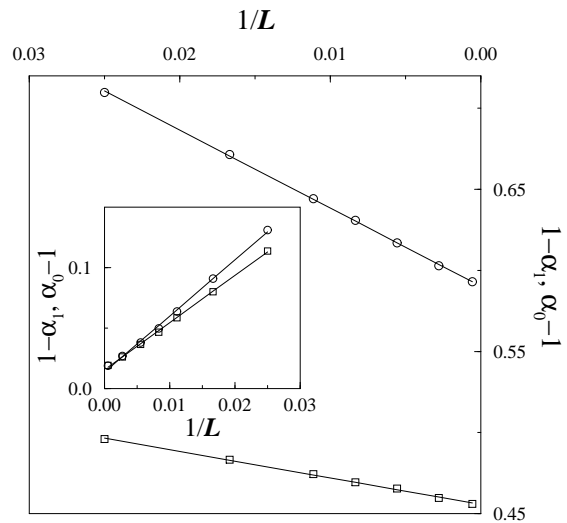}
\caption{\label{fig2} The $L$ dependence of the scaling exponents
$\alpha_0$ (circles) and $\alpha_1$ (squares) for the 1D disordered
system in the strong-coupling regime ($b=0.3)$. The straight lines
are linear fits to Eq. (\ref{s2l}). The inset shows the same dependence
for the case of weak coupling ($b=10$).}
\end{center}
\end{figure}

In Fig. 2, we represent the finite-size corrections for the
scaling exponents $\alpha_0$ (circles) and $\alpha_1$ (squares)
for the 1D disordered system described by Eq. (\ref{h1dor}) in the
strong-coupling regime ($b=0.3)$. The straight lines are linear
fits to Eq. (\ref{s2l}) and intercept the vertical axis at $0.590$
and $0.455$, respectively. These results clearly demonstrate that
Eq. (\ref{a0a1}) is not fulfilled at all, indicating that the PS
is no longer valid in this case. This is not surprising since
the parameter found ($\epsilon=0.590$), is not small, and the
$\epsilon$ expansion (\ref{parap}) can only be justified
parametrically for $\epsilon \ll 1$. We have found that if $b$
is reduced, the intercept points shift to higher values and their
separation increases.

The inset of Fig. 2 shows the same $L$ dependence for the case of
weak coupling ($b=10$). Unlike in the strong-coupling regime, both
lines intercept the $y$ axis practically at the same values, of
$0.0194$ and $0.0193$, respectively. This suggests that $f(\alpha)$ is
parabolic, in agreement with Eq. (\ref{parap}). In this case,
$\epsilon=0.0194 \ll 1$, as expected. Very recently, results similar
to those in the inset were obtained for the Chalker-Coddington
model, although in this case a different method \cite{EM01} was used
to evaluate $\alpha_q(L)$. In this calculation, the parameter found
was $\epsilon=0.26$.
\begin{figure}
\begin{center}
\includegraphics[width=8.0cm]{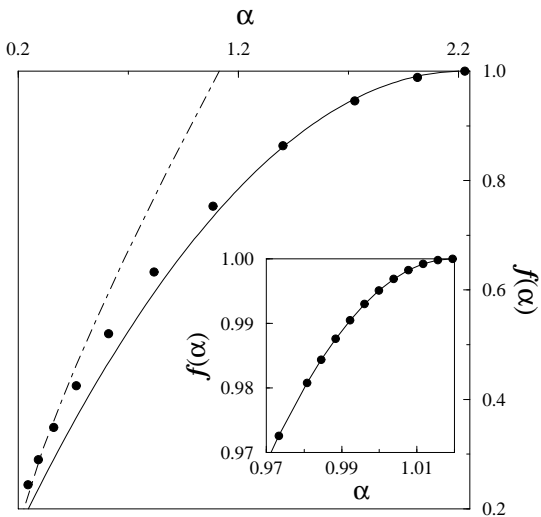}
\caption{\label{fig3} The $f(\alpha)$ spectrum (solid circles) for
the 1D disordered system in the strong-coupling regime ($b=0.1$).
The implicit parameter $q$ ranges from 0 to 1. The PS with
$\alpha_0=2.228$ is shown for comparison (solid line). The dot-dashed
line corresponds to Eq. (\ref{fdea}) for $q> 1/2$. The inset shows the
same spectrum in the case of weak-coupling ($b=10$) together with the
PS with $\alpha_0=1.0194$ (solid line).}
\end{center}
\end{figure}

The extrapolated results of the whole $f(\alpha)$ curve in the
strong-coupling regime $b=0.1$ are depicted in Fig. 3 (solid circles).
The implicit parameter $q$ ranges from 0 to 1, in steps of 0.1. Note that
Eq. (\ref{parap}) is assumed to be valid least up to $|q| \le 1$. \cite{Ja94}
The PS with $\alpha_0=2.228$ is shown for comparison (solid line). The
difference between both results is noteworthy. Based on a resonance approximation, 
the following formula was derived in  Ref. 16 for $f(\alpha)$ in the regime
$b \ll 1$ for values of $q>\tfrac{1}{2}$,
\begin{equation}
f(\alpha)=2bF(\alpha/2b)\;,\label{fdea}
\end{equation}
where $F(A)$ is the Legendre transform of the function
$\Gamma(2q-1)/2^{2q-3}\Gamma(q)\Gamma(q-1)$. The result of Eq. (\ref{fdea})
for $b=0.1$ is represented by a dot-dashed line. At $q=\tfrac{1}{2}$, the
mass exponent $\tau_q$ diverges, while
for smaller values of $q$, this approximation completely breaks down.
Note that $f(\alpha)$ properly interpolates between the parabola and
Eq. (\ref{fdea}). For the smallest values of $q$ near the maximum,
$f(\alpha)$ is very close to the PS, and as $q$ increases, $f(\alpha)$
separates from this curve with a regular tendency until it reaches
the resonance prediction for the larger values of $q$ as shown.
We have found that as $b$ decreases, the deviations with respect to PS
are larger, and as $b$ increases, these deviations diminish. For values of
$b$ well inside the weak-coupling domain (see inset), the deviations are
practically unobservable.
The inset of Fig. 3 shows the same $f(\alpha)$ spectrum for the weak
coupling regime ($b=10$). The perfect agreement between the obtained
results (solid circles) and the PS, Eq. (\ref{parap}), with
$\alpha_0=1.0194$ is evident. Similar results were obtained in
Ref. 16 for this regime. We can thus conclude, as we previously
stated, that the PS is correct in this case.

\subsection{2D system}

We have also calculated the $f(\alpha)$ spectrum in the
experimentally more important case $d=2$. Unlike the 1D PRBM model,
until now, it has not been possible to solve the 2D disordered models with
long-range transfer terms analytically. We consider noninteracting
electrons on a 2D square lattice with random on-site potentials
$\varepsilon_i$ and random transfer terms
$V_{ij}=V\varepsilon_{ij}/|\bm{r}_i-\bm{r}_j|^2$, whose Hamiltonian reads
\begin{equation}
{\cal H}=\sum_i\varepsilon_i|\bm{r}_i\rangle\langle \bm{r}_i|+
\sum_{i\ne j} V_{ij} 
|\bm{r}_i\rangle\langle \bm{r}_j|\;,
\label{Model}
\end{equation}
where the vectors $\bm{r}_i$ label the sites of the lattice, and
$\{\varepsilon_i\}$ and $\{\varepsilon_{ij}\}$ are two sets of
uncorrelated random numbers uniformly distributed within the interval
$(-W/2, W/2)$, and $(-S/2, S/2)$, respectively. $V=1$ defines the
energy scale and $S/V$ is taken to be equal to $1$ in all regimes. Each
regime is characterized by the coupling strength $\lambda=W/S$ and we
take $0.4 \le W/V \le 30$. The system sizes used are $L=24$, 36, 56,
72, and 96. The critical properties of this model in the strong disorder
regime have recently been investigated numerically. \cite{PS02}
A closely related model in $d=1-3$ was also considered in
Ref. 25 for strong disorder.

In Fig. 4, we show the same size corrections as in Fig. 2 for the 2D
disordered system in the strong-coupling regime ($\lambda=6$). The
fitted straight lines intercept the vertical axis at $0.827$ and $0.743$.
The corresponding results for the case of weak coupling ($\lambda=0.4$) are
reported in the inset. As in the 1D system, we obtained a $L^{-1}$ behavior
for these corrections in both regimes. All the comments made with respect
to the 1D system are equally valid in this case. Hence, we arrive at the
same conclusions for the 2D case, i.e., the PS is valid in the weak
disorder regime whereas the strong disorder case strongly deviates from
this behavior.
\begin{figure}
\begin{center}
\includegraphics[width=8.0cm]{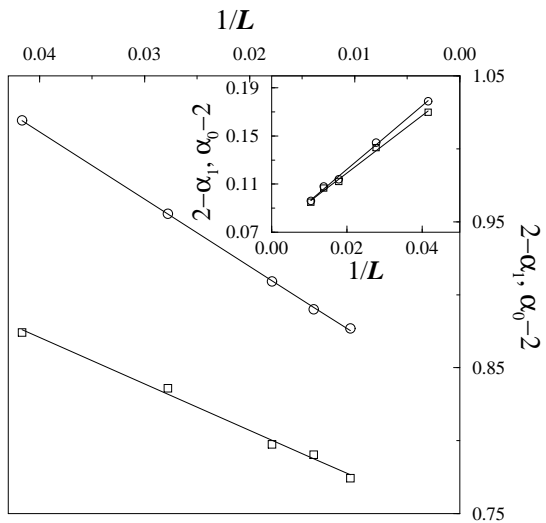}
\caption{\label{fig4} As for Fig. 2, for the 2D disordered system
in the strong-coupling regime ($\lambda=6$). The results of the
inset correspond to the same system in the weak-coupling regime
($\lambda=0.4$).}
\end{center}
\end{figure}
\begin{figure}
\begin{center}
\includegraphics[width=8.0cm]{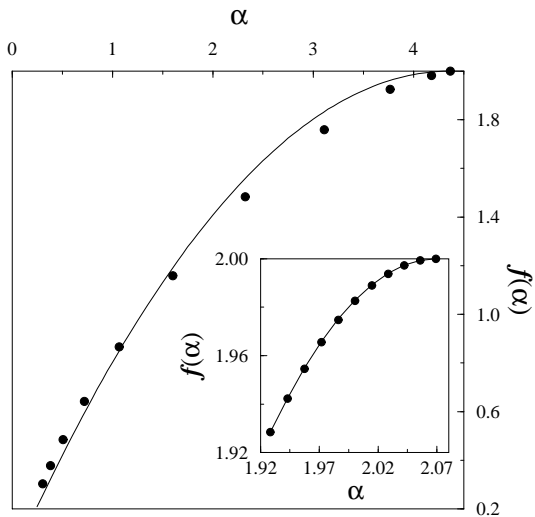}
\caption{\label{fig5} Same as Fig. 3, for the 2D disordered
system in the strong-coupling regime ($\lambda=30$). The PS
with $\alpha_0=4.3673$ is shown for comparison (solid line).
The inset shows the same spectrum in the case of weak coupling
($\lambda=0.4$) together with the PS with $\alpha_0=2.0692$
(solid line).}
\end{center}
\end{figure}

Figure 5 shows the same extrapolated values of the $f(\alpha)$ spectrum
as in Fig. 3 for the 2D disordered system in the strong disorder regime
($\lambda=30$). The PS with $\alpha_0=4.3673$ is shown for comparison
(solid line). As in the 1D case, the difference between both results is
notable. As there are no analytical predictions for this regime, we are
not able to justify the sign of the deviations in this case. The inset
shows the same $f(\alpha)$ spectrum (solid circles) for the weak-coupling
regime ($\lambda=0.4$) together with the PS (solid line) with
$\alpha_0=2.0692$.
\begin{figure}
\begin{center}
\includegraphics[width=8.0cm]{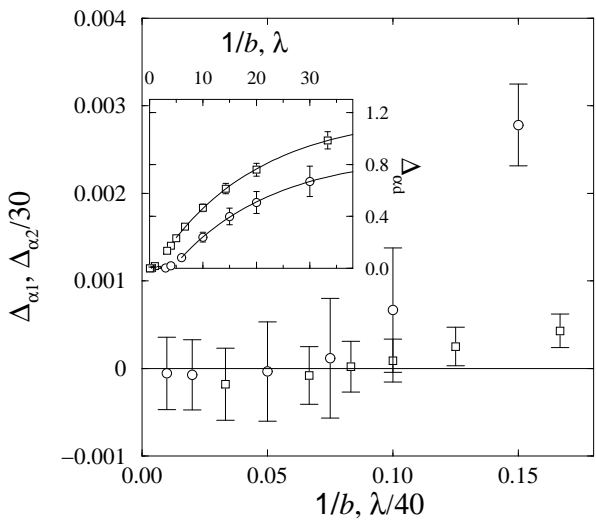}
\caption{\label{fig6} The small coupling constant dependence of the
parameter $\Delta_{\alpha d}$ for the 1D (squares) and 2D (circles)
disordered systems. The inset shows the same dependence in the whole
range of $b$ and $\lambda$ studied; solid lines are fits  to Eq.
(\ref{delta}).}
\end{center}
\end{figure}

We now define in a simple way the extent to which the spectra deviate from
parabolicity. We consider the distance between the two intercept points
in Figs. 2 and 4, $\Delta_{\alpha d}=\alpha_0+\alpha_1-2d$, as a measure
of departure from the parabolic shape. This clearly distinguishes a
parabolic spectrum from a nonparabolic one. In the following, we will use
the notations $t_1=1/b$ and $t_2=\lambda$ for the coupling constant
of the 1D and 2D models, respectively. In Fig. 6, we represent the dependence
with the coupling constant $t_d$ of the parameter $\Delta_{\alpha d}$ for
1D (squares) and 2D (circles) disordered systems. This
parameter will tend to a well-defined finite value in the limit
$t_d \to \infty$. The small $t_d$-dependent behavior of $\Delta_{\alpha d}$,
corresponding to the weak-coupling regime, is shown in the main panel of Fig. 6,
in which one can clearly appreciate that below a given value of $t_d$
($t_1 \approx 0.1$ for the 1D system and $t_2 \approx 4$ for the 2D
case), $\Delta_{\alpha d}=0$ within the error bars. This result strongly
suggests that all corrections to parabolicity vanish at a \textit{finite}
value of the coupling constant, in contrast to what might be expected naively.

In order to predict the large $t_d$ asymptotic values of
$\Delta_{\alpha d}$ from the inset of Fig. 6, a curve of the form
$\Delta_{\alpha d}=\Delta_{0}-G_{A,B}(t_d)$ was fitted to the
data points in this  graph,  where $\Delta_{0}$, $A$, and $B$ are three
positive fitting parameters, and $G_{A,B}(t_d)$ is a function
which goes to zero as $t_d \to \infty$. Note that this fit was
performed only over the larger values of $t_d$ where convergence
was evident. Various forms for $G_{A,B}(t_d)$ were chosen and
tested on the data plots. These included exponential, inverse logarithmic,
and power-law decay with $t_d$. The testing involved performing a
three-parameters fit on the $\Delta_{\alpha d}$ versus $t_d$ data plots
using the standard Levenberg-Marquardt method for nonlinear fits. The form
of $G_{A,B}(t_d)$ eventually chosen was the exponential law.
The reason for choosing the above form of $G_{A,B}(t_d)$ with preference to
the alternative forms was simply that the inverse logarithmic and the power
laws do not fit our data properly. Thus, the small and large $t_d$-dependent
behavior of $\Delta_{\alpha d}$ can be described by
\begin{equation}
\Delta_{\alpha d}=
\left\{\begin{array}{ll}
       0 \quad \quad \quad \quad \quad ,  & t_1 \alt 0.1, \;t_2 \alt 4\\
       \Delta_0-Ae^{-Bt_d} \,\ ,\quad & t_d \gg 1 \;.
       \end{array}\right.
\label{delta}
\end{equation}
\noindent
The solid lines in the inset of Fig. 6 are fits to Eq. (\ref{delta}) with
parameters $\Delta_0=1.19 \pm 0.02$, $A=1.250 \pm 0.013$, and
$B=0.054 \pm 0.002$ for the 1D system, and $\Delta_0=0.88 \pm 0.02$,
$A=1.110 \pm 0.013$, and $B=0.055 \pm 0.002$ for the 2D case.

Very recently, Potempa and Schweitzer \cite{PS02} reported the $f(\alpha)$
spectrum for a critical wave function of a finite sample in 2D of linear
size $L=150$ for $\lambda=6$. For this value of $\lambda$, we found
$\Delta_{\alpha2}=0.084$, indicating that we are very close to the regime
where deviations are practically unobservable. These authors found
deviations (downwards) from PS mainly for the largest negative $q$ values.
We believe that these deviations are due to the use of the finite-size value
of $\alpha_0$. If the extrapolated value of $\alpha_0=2.827$ is used instead,
the agreement with the PS is fairly good.

Before concluding, we must admit that experiments to measure $\alpha_0$
or related multifractal exponents have not been carried out. However,
the statistical properties of wave functions are close related
to the conductance statistics. As was demonstrated in Ref. 9,
multifractality can show up in transport experiments.
Specifically, the $q$th moments of the two-point conductance exhibit
multifractal statistics. The connection between the PS for the
conductance and Eq. (\ref{parap}) is through the relation
$X_t=2(\alpha_0-d)$, where $X_t$ is the power-law exponent of the
typical conductance. In this way, our results could be checked in
experiments with random systems using samples with varying disorder.

\section{Conclusions}

In this paper, we have calculated the multifractal scaling exponents
of the wave functions for 1D and 2D disordered systems with long-range
transfer terms at criticality.
The leading finite-size corrections to $\alpha$ and $f(\alpha)$ decay
algebraically with exponents equal to $-1$.
In the limit $L \to \infty$, we have 
demonstrate that according to theoretical predictions, the $f(\alpha)$
spectrum is parabolic in the  weak disorder regime. Our
calculations strongly suggest that the parabolic spectrum completely
breaks down for large values of the disorder strength, and more
importantly, that all corrections to the parabolic shape vanish
(within our numerical uncertainties) at a finite value of the coupling
constant.

The question arises as to whether these results are applicable to
other quantum systems, particularly in the 3D Anderson transition
that occurs in the strong disorder domain, and whose similarity with
the PRBM model at $b=0.3$ has been demonstrated for several critical
magnitudes. \cite{CO02} Another interesting question is whether the
deviations studied here also have signatures in other properties, such
as spectral statistics or generalized dimensions.
Finally, how the present 2D model can be studied, using 2D conformal
field theory or locator expansion instead of the resonance pair
approximation, remains to be resolved.

\begin{acknowledgments}
The author thanks the Spanish DGESIC for financial support through
Project No. BFM2000-1059.
\end{acknowledgments}


\begin{thebibliography}{}

\bibitem{FE95} V.I. Falko and K.B. Efetov,
Europhys. Lett. \textbf{32}, 627 (1995);
Phys. Rev. B \textbf{52}, 17413 (1995).

\bibitem{Ja94} M. Janssen,
Int. J. Mod. Phys. B \textbf{8}, 943 (1994);
B. Huckestein, Rev. Mod. Phys. \textbf{67}, 357 (1995).

\bibitem{Ef83} K.B. Efetov,
Adv. Phys. \textbf{32}, 53 (1983).

\bibitem{CP86} C. Castellani and L. Peliti,
J. Phys. A \textbf{19}, L429 (1986);
F. Wegner, Nucl. Phys. B \textbf{280}, 210 (1987).

\bibitem{MC96} C.C. Chamon, C. Mudry, and X.-G. Wen,
Phys. Rev. Lett. \textbf{77}, 4194 (1996);
H.E. Castillo, C.C. Chamon, E. Fradkin, P.M. Goldbart,
and C. Mudry, Phys. Rev. B \textbf{56}, 10668 (1997).

\bibitem{CD01} David Carpentier and Pierre Le Doussal,
Phys. Rev. E \textbf{63}, 026110 (2001).

\bibitem{BK00} M.J. Bhaseen, I.I. Kogan, O.A. Soloviev,
N. Taniguchi, and A.M. Tsvelik, Nucl. Phys. B \textbf{580},
688 (2000).

\bibitem{Zi99} M.R. Zirnbauer, hep-th/9905054 (unpublished).

\bibitem{JM99} M. Janssen, M. Metzler and M.R. Zirnbauer,
Phys. Rev. B \textbf{59}, 15836 (1999).

\bibitem{CJ89} Ashvin Chhabra and Roderick V. Jensen,
Phys. Rev. Lett. \textbf{62}, 1327 (1989).

\bibitem{An58} P.W. Anderson,
Phys. Rev. \textbf{109}, 1492 (1958).

\bibitem{Le89} L.S. Levitov,
Europhys. Lett. \textbf{9}, 83 (1989);
Phys. Rev. Lett. \textbf{64}, 547 (1990).

\bibitem{Le99} L.S. Levitov,
Ann. Phys. (Leipzig) \textbf{8}, 697 (1999).

\bibitem{Cu99} E. Cuevas, Phys. Rev. Lett. \textbf{83},
140 (1999); E. Cuevas, E. Louis, and J.A. Verg\'es,
\textit{ibid}. {\bf 77}, 1970 (1996).

\bibitem{Mi00} A.D. Mirlin, 
Phys. Rep. \textbf{326}, 259 (2000);
F. Evers and A.D. Mirlin,
Phys. Rev. Lett. \textbf{84}, 3690 (2000).

\bibitem{ME00} A.D. Mirlin and F. Evers,
Phys. Rev. B \textbf{62}, 7920 (2000).

\bibitem{CG01} E. Cuevas, V. Gasparian and M. Ortu\~no,
Phys. Rev. Lett. \textbf{87}, 056601 (2001);
E. Cuevas, Phys. Rev. B \textbf{66}, 233103 (2002).

\bibitem{CO02} E. Cuevas, M. Ortu\~no, V. Gasparian,
and A. P\'erez-Garrido, Phys. Rev. Lett. \textbf{88},
016401 (2002).

\bibitem{Va02} I. Varga and D. Braun,
Phys. Rev. B \textbf{61}, R11859 (2000);
Imre Varga, \textit{ibid}. \textbf{66}, 094201 (2002). 

\bibitem{KM97} V.E. Kravtsov and K.A. Muttalib,
Phys. Rev. Lett. \textbf{79}, 1913 (1997).

\bibitem{MF96} A.D. Mirlin, Y.V. Fyodorov, F.M. Dittes,
J. Quezada, and T.H. Seligman,
Phys. Rev. E \textbf{54}, 3221 (1996).

\bibitem{KT00} V.E. Kravtsov and A.M. Tsvelik,
Phys. Rev. B \textbf{62}, 9888 (2000).

\bibitem{EM01} F. Evers, A. Mildenberger, and A.D. Mirlin,
Phys. Rev. B \textbf{64}, 241303(R) (2001).

\bibitem{PS02} H. Potempa and L. Schweitzer,
Phys. Rev. B \textbf{65}, 201105(R) (2002).

\bibitem{PS98} D. A. Parshin and H. R. Schober,
Phys. Rev. B \textbf{57}, 10232 (1998).


\end{thebibliography}
\end{document}